\documentclass[11pt]{journal}
\usepackage[english]{babel}
\usepackage{setspace} % For setting double spacing
\usepackage[letterpaper]{geometry}
\usepackage{longtable}
\usepackage{booktabs}
\usepackage{amsmath}
\usepackage{times}
\usepackage[colorlinks=true, linkcolor=blue,  citecolor=blue, urlcolor=blue]{hyperref}
\usepackage{float}
\usepackage{graphicx}
\usepackage{times}
\usepackage[authoryear,round]{natbib}

\usepackage{fancyhdr}
\title{What does making money have to do with crime?: A dive into the National Crime Victimization survey}

\author{% 
  Sydney Anuyah \thanks{%
    Luddy School of Informatics, Engineering and Computing\\
    \texttt{sanuyah@iu.edu}%
  }%
}
\begin{document}
\maketitle

\section*{Abstract}
In this short article, I leverage the National Crime Victimization Survey (NCVS) from 1992 to 2022 to examine how income, education, employment, and key demographic factors shape the type of crime victims experience (violent vs. property). Using balanced classification splits and logistic regression models evaluated by F1-score, there is an isolation of the socio-economic drivers of victimization (``Group A" models), and then an introduction of demographic factors such as: age, gender, race, and marital status controls (``Group B" models). The results consistently proves that higher income and education lower the odds of violent relative to property crime, while men, younger individuals, and racial minorities face disproportionately higher violent-crime risks. On the geographic spectrum, suburban models achieve the strongest predictive performance (Accuracy 0.607, F1 0.590), urban areas benefit from adding education and employment predictors, and crime in rural areas are still unpredictable using these current factors. The patterns found in this study shows the need for specific interventions like educational investments in metropolitan settings, economic support in rural communities, and demographic-aware prevention strategies.

\section{Introduction}
\subsection{Background}
Criminologists have long debated the causes of crime, producing a wide array of theories to explain why crimes happen. Despite the substantial progress in this field, no single model has successfully captured the full complexity of criminal behavior. Various approaches, such as biological theories \citep{raine2002biological}, deterrence theories \citep{tomlinson2016examination}, psychological theories \citep{moore2011psychological}, and other models, have tried to explain different aspects of crime. However, each theory has limitations in its scope, as these theories usually address only specific aspects of criminal behavior. Therefore, we often observe correlation at best. 

Using an integrated approach—where multiple theories are grouped together—has arguably been more effective than relying on single-theory explanations \citep{krohn2014integrated}. This is because it combines evidence from multiple perspectives to provide a more holistic understanding of crime. The challenge of this approach, as highlighted in previous research, lies in managing theoretical complexity and reconciling conflicting evidence \citep{krohn2014integrated}. 

To test these theories for validity, criminologists need robust datasets to identify correlations or causal relationships. No theory can be considered valid without sufficient empirical evidence to support its claims. Therefore, data is essential in understanding crime and crime victimization.

It is assumed that for any crime to occur, three elements must be present: (1) the assailant, (2) the victim, and (3) an enabling environment. Criminologists typically categorize crimes into two broad types: property crime and violent crime \citep{abarbanel2001victim}. Property crime involves the theft or destruction of someone's property without the use or threat of force against the victim. Violent crime, on the other hand, involves the use or threat of physical force intended to harm or intimidate another individual. 

In this study, we will examine these two major crime categories.

It is however good to note that, some crimes  are described as victimless crimes (e.g. prostitution, embezzlement, etc)\citep{roseblasphemy}. In these cases, there is a search for the victim, or the assailant might also be the victim. This is what makes having a one-size fit all definition practically difficult. Crime victimization which is defined as the experience of being impacted as a result of a criminal act \citep{morgan2019criminal} is a key but often under-reported phenomenon. Common crime reporting methods, such as police records, do not always accurately capture the full scope of criminal activity, especially unreported crimes. The National Crime Victimization Survey (NCVS), introduced in the 1970s, was designed to fill this gap by collecting data directly from victims. By capturing victims of crimes especially those that go unreported to law enforcement, the NCVS is believed to offer a more accurate and holistic picture of crime trends\citep{roman2022causal} analyzed the NCVS dataset and explains how it has shed light on some changes in societal crime patterns and informed more effective policymaking.

\subsection{Problem Statement}
In this study, the author plans to use the NCVS data to understand how the amount of money a person makes given their work location directly has an impact on what kind of crime they are susceptible to (property or violent crime), given that these individuals are already victims of crime. In looking at income, we have to consider other socio-economic factors that play an important role in affecting income levels. In the dataset, we have Education and Employment as additional socioeconomic factors that primarily drive income levels. Our first hypothesis from our socioeconomic variable tells us that if a person is indeed victimized, the probability of the crime being a violent crime when compared to a property crime reduces as the income gets larger.

Furthermore, socio-economic factors are strongly tied with demographics and we need to control for these factors to have a balanced approach \citep{worthen2024metoo}. Here, factors such as Age,  Marital status, Gender, and Race are considered because they are both very important datasets and present in the dataset. We also hypothesize that age is influential in understanding violent crime victims. It is assumed that younger individuals are more prone to violent crimes due to mobility, social interactions, and peer influence. This will also be tested through our analysis. Furthermore, our third hypothesis will be that men are more likely to be victims of violent crime when compared to property crime than women. 

For a more robust analysis, we will look at the geographical context. In urban areas, characterized by high population densities, high demand for employment, income and eduction, with diverse demographic conditions, it usually exhibit a higher elevated crime rates compared to suburban and rural areas. Now, diving deep into the subject, we want to know if in these regions, we see in a certain region, a lean towards one certain type of crime to another, and understand what factors affect the type of crime on a granular level. 

The reason for choosing the NCVS dataset for this analysis was because of its comprehensive nature in collecting crime data especially for un-reported incidents. The timeline spanning from 1992 to 2022 is chosen to ensure that the analysis encompasses a substantial period that captures both historical and contemporary crime trends. This 30-year span allows for the examination of long-term patterns and the assessment of how socio-economic shifts, policy changes, and societal developments have influenced victimization rates over time

\subsubsection{Research Relevance and Questions}
Understanding the why and how variables interact the way they do will assist in understanding crime victims and eventually, create policies to reduce the crime rate based on analysis. Overall, the goal is to build a solid understanding of how income influences violent crimes vs property crimes and how other factors which affect the income level might play a significant role in shaping the final outcome. This  research aims to answer the following questions: 

\begin{enumerate}
    \item If we know a person currently has suffered crime victimization, can we predict from their income and other socioeconomic and demographic factors what kind of crime it was?
    \item How does the work location (rural vs. urban vs. suburban) influence the likelihood of crime victimization?
    \item What is the relationship between controlling for demographic factors and not controlling for them when using socioeconomic factors to understand crime victimization?
    \item Can income really be a good predictor of crime victimization?
\end{enumerate}

\section{Literature Review}
\subsection{Victimization patterns across socio-economic factors}
As evidenced by NCVS data, the level of crime and the likelihood of being a victim of crime are strongly related to certain socioeconomic characteristics, including level of education, employment status, and income. It is noted by \citep{roseblasphemy} that the crime rates are susceptible to the overall economic structural changes including level of unemployment or the extent of earnings dispersion. It has been found that those who attain higher levels of education possess lower rates of both committing crimes and being victimized, indicating that education plays a protective role through higher chances of finding stable jobs and lower economic pressure. On the other hand, people with lower levels of education, who have higher chances of residing in urban areas are characterized by poverty and low social cohesion, are at a higher likelihood of being crime victims. The NCVS data also reports that the unemployed or those who hold only marginally regular and dangerous jobs have disproportionate chances of being victimized, especially in terms of property crime \citep{ncvs2023}. This is consistent with the routine activity theory of crime, which suggests that those who may have no regular work hours may be more susceptible to certain risky conditions. Other factors responsible for the fall in burglary rates since 1970 are also described by \citep{farrell2022forty} as the increase in employment levels and economic stability as well as the introduction of security advancements. Furthermore, ties of income inequality deepen the chances of being victimized. According to \citep{xie2021immigrant} members of the low income group have a greater chance of being victimized. However, they do not report many crimes due to mistrust of law enforcement agencies and belief that justice will not be served\citep{worthen2024metoo} point out that movements like \#MeToo should also change the situation because they enable victims from different economic and social backgrounds to come forward with their crimes, such as sexual violence.

\subsection{Victimization patterns across demographic factors}
Critical demographic factors determining the pattern of victimization and the character of the crime are age, marital status, race, and gender. Of these, age is the most influential, with younger people, particularly those between the ages 18 and 24, being the most vulnerable to victimization \citep{hayes2022heterogeneity} Taking Note of the age group, here, \citep{hayes2022heterogeneity} explain that this younger group has even more cultural exposure to risk, such as nightlife socialization activities. In contrast, the older population, while not as frequent victims of crime, have further complications such as the risk of being a target of financial fraud more often. Also, older crime victims might refrain from reporting incidences of crime to the authorities because of a feeling of shame or fear of being dependent and losing their self-strength, hence why the NCVS and similar datasets would make victimization more opaque \citep{ncvs2023}. The victimization trends are also well explained with the marriage status of the victims. A commonly repeated finding is that unmarried and divorced victims are those who are at relatively greater risk of being victimized due to their higher chances of living alone or of being more active outside the home where chances of exposure to crime are higher \citep{USCCR2024}. Engaging in unsupervised activities in high-risk environments is less common among married individuals, so there is less need for extensive protective measures. However, marriage can act in the opposite direction when it comes to domestic violence reporting as a large number of victims of such violence face many constraints such as fear of retribution or being economically dependent on the perpetrator \citep{truman2024characteristics}. For some specific groups, particularly, women and racial minorities, who may experience racism and sexual violence, race and gender factors entangle to increase chances of victimization. According to research by \citep{flores2023violent}, the violent crime victimization levels of racial minorities, most octantly black and Hispanic persons, are considerably higher, often associated with poverty and dislocation. More gender disparities exist for women who are at a higher chance of facing sexual and domestic violence\citep{worthen2024metoo}, for instance, \citep{worthen2024metoo} highlights the impact of the social movement \#MeToo on the increase of complaints about such crimes, especially ones that go unreported. For example,  men are less prone to sexual violence, but are more prone to violence in general, particularly in public places. 

\section{Methodology}
\subsection{NCVS Design, Crime Coverage, and Victim Reporting Patterns}
The NCVS data \citep{ncvs2022} is representative of the US population. When conducting its research to create this dataset, the NCVS employed a multi-stage stratified sampling technique where various counties, regions and a variety of households are included. This strategy enabled the study to take on board a large diversity of victimization and therefore generalize its solutions across subpopulation strata \citep{ncvs2023}. This notwithstanding, \citep{xie2024declining} are critical of this approach and methodology as a whole – that even with its robust design and methodology, sensitive groups like the homeless or transient workers tend to be excluded – widening the net comprehensiveness of the survey. For the NCVS, data quality control and credibility is very critical for its effectiveness. While their methodology minimizes sampling bias, other factors like those from recall bias and response fatigue are likely to affect the findings\citep{langton2023report}. Recall bias particularly contributes to under-reporting of low level and postponed minor crimes or even the timeline events, and response fatigue could cause lower data consistency during the last stages of the interview. On the other hand, studies like \citep{pennayresults} seem to support the NCVS for they regard the latter as a gold standard in the collection of crime figures. This is because of the elaborate design of the NCVS – supplemented by far by the large scale it covers.

The NCVS encompasses a three major crime categories which in turn encompasses multiple crimes. These are violent crimes (e.g. assault, robbery, rape, etc.), \citep {blondel2024crimes}, property crimes (e.g. burglary, larceny and motor vehicle theft) and a combination of both. This distinction assists researchers in appreciating the different types of crimes and their pronounced communities. For example, the emotional and physical effects of violent crimes are generally severe, while property crimes can be classified as less severe since the outcome is mostly financial loss. In such a manner, by differentiating among kinds of victimization, the NCVS assists in informing the policy makers in addressing the problem of victimization. However, it should be noted that while the NCVS has a wide scope, it unfortunately excludes some forms of crime altogether like the victimless crimes of this kind as hatespeech, drug addiction or these that fall under white collar crimes such as embezzlement and frauds\citep{morgan2022nation}. Similarly, stalking, identity theft, cybercrime are also not included as they pose their own set of complications. Although the NCVS has begun adjusting its parameters in order to contain such offences, \citep{holt2024assessment} emphasizes that its methods are not appropriate for the digital era. For example, stalking’s intrusiveness outpaces reports of its occurrence because many victims are unaware of the crime being perpetrated against them.

An additional merit of the NCVS is that it permits both the reporting of crime and the research into the possibility of crime that has not been reported, providing an in-depth interpretation of victimization other than police records. Unfortunately, several forms of harm or violence, especially domestic violence or petty theft or vandalism, are seldom reported to the authorities. The NCVS is able to pinpoint these areas, and show how many crimes do go unregistered within conventional administrative jurisdictions \citep{thompson2022criminal}. The reporting behavior patterns differ greatly depending on the sex and socio-economic status. For instance, victims of violent crimes are more likely to report than those of non-violent crimes, as the former’s severity is usually considered more serious \citep{harvey2022reducing}. Other than that, factors such as age and race influence behavior in the reporting process, as they are defined as demographic factors. Women, for example, may be reluctant to contact authorities about sexual violence and threat because of social pressure, whereas minorities, in particular race-based minorities, face structural obstacles that impede their willingness to report to police \citep{xie2021immigrant}. This is why we want to study how income, a socio-economic factor, interplays with other socioeconomic factors and how these also interplay with given demographic factor given the work-location.

\subsection{Data Preparation}
NCVS is publicly available as a concatenated file from 1992-2022 spanning 30 years, on the ICPSR website \citep{ncvs2022}. For a quicker analysis, it was easier to use Python programming language for such a large file than using statistical software like STATA. The dataset contained 6,030,292 rows of which 98\% of that value contained missing information. Since we were focused on the following variables: Income, Age, Marital Status, Gender, Race, Education, Employment, Work Location and Victimization Type (which is our only dependent variable), it was intuitive to filter out the rows that had missing values in any of these cases, and we were left with 139,648 rows representing 2.3\% of our dataset. The subsections detail how the categorical variables were coded. Variables like age that were numerical were used directly. Age ranged from 16-90 years. For all categorical variables, the \textbf{Code 0} was the one that was dummied variable for the analysis.

\subsubsection{Preparing Income}
The codes for income were extrapolated directly from the NCVS and were recoded here as \textbf{Code 0:} representing people earning \textit{Less than \$5,000}.  \textbf{Code 1:} representing household income between \textit{\$5,000 -\$7,499},  \textbf{Code 2:} representing  \textit{\$7,500- \$9,999},  \textbf{Code 3} representing \textit{\$10,000 - \$12,499}, \textbf{Code 4} representing \textit{\$12,500 - \$14,999}, \textbf{Code 5} representing \textit{\$15,000 - \$17,499},  \textbf{Code 6} representing \textit{\$17,500 - \$19,999}, \textbf{Code 7} representing \textit{\$20,000 - \$24,999}, \textbf{Code 8} representing \textit{\$25,000 - \$29,999}, \textbf{Code 9} representing \textit{\$30,000 - \$34,999}, \textbf{Code 10} representing \textit{\$35,000 - \$39,999}, \textbf{Code 11} representing \textit{\$40,000 - \$49,999}, \textbf{Code 12} representing \textit{\$50,000 - \$74,999}. Income levels of \textit{\$75,000 and \$99,999} are represented by \textbf{Code 13}. After the first filtration, there were no income values exceeding the \$99,999 for our use case. Residue values were deleted. 

\subsubsection{Preparing Marital Status }
In marital status the following are the codes: \textbf{Code 0:} representing  \textit{married}.  \textbf{Code 1:} representing \textit{widowed},  \textbf{Code 2:} representing  \textit{divorced},  \textbf{Code 3} representing \textit{separated}, \textbf{Code 4} representing \textit{Never Married}.

\subsubsection{Preparing Gender}
In gender the following are the codes: \textbf{Code 0:} representing  \textit{male} and  \textbf{Code 1:} representing \textit{female}. The first filter removed all other entries of other genders, and over 78.6\% had missing data, and 4.8\% was residue. Therefore, these were eliminated.

\subsubsection{Race}
For racial identites the following are the codes: \textbf{Code 0:} representing  \textit{White}.  \textbf{Code 1:} representing \textit{Black},  \textbf{Code 2:} representing  \textit{American Indian, Aleut, Eskimo},  \textbf{Code 3} representing \textit{Asian, Pacific Islander}, \textbf{Code 4} representing \textit{Other}.

\subsubsection{Education}
Education represents the highest level of education attained as of the time of the interview. \textbf{Code 0:} representing the \textit{Kindergarten or Never been to school}, \textbf{Code 1-12} representing the \textit{12 grades of school from Elementary to High School}. \textbf{Code 21-26} representing the \textit{1, 2, 3, 4 5 and 6 years of education in college, excluding graduate degrees}. Other codes were filtered out because of incomplete data.

\subsubsection{Employment}
The Employment codes are: \textbf{Code 0:} representing  \textit{A private company, business, or individual for wages}.  \textbf{Code 1:} representing \textit{Federal government},  \textbf{Code 2:} representing  \textit{A state, county, or local government},  \textbf{Code 3} representing \textit{Yourself, (self-employed) in your own business, professional practice, or farm}. Code 3 will also cover the population who are categorized as unemployed.

\subsubsection{Work Location}
The work location codes are: \textbf{Code 0:} representing  \textit{Urban area}.  \textbf{Code 1:} representing \textit{Suburban area},  \textbf{Code 2:} representing  \textit{Rural area},  \textbf{Code 3} representing \textit{Combination of any of these}.

\subsubsection{Victimization Type}
The victimization type had just two codes here, as we are just comparing property and violent crimes.  \textbf{Code 0:} representing  \textit{violent crimes}.  \textbf{Code 1:} representing \textit{property crimes}.

\subsection{Finding the Best Fit for the Model}
Due to the large data imbalance between violent crimes at 18.18\% and property crimes at 81.82\%, it was very easy for the model to directly predict that all crimes are property crimes and it would be correct 82 out of 100 times. However, this approach would not be meaningful for us to make any useful comparisons. This led us to explore the F1-score, which is the harmonic mean comprising of precision and recall. Precision is the ratio of true positive predictions to the total number of positive predictions made by the model, indicating the accuracy of the positive class predictions. Recall, on the other hand, is the ratio of true positive predictions to the actual number of positive instances in the dataset, reflecting the model's ability to identify all relevant cases.

Using the F1 score  enabled us to achieve a balance between the recall and the precision of the model which in turn manages the class imbalance problem in the identification of both property and violent crimes. A high F1 score suggests that the model is able to predict the majority class and make fewer incorrect predictions.  The way this was done was different iterations of the dataset class sizes were tested and the best iteration of a model with the highest F1-score was chosen. Eventually, the data size that had the highest F1-score was a 50:50 split. To ensure a pure 50:50 split, we randomly selected an equal number of property crime variables to build the model. The seed for reproducibility was set to 564.

\section{Results and Analysis}

\subsection{Group A Models - Socioeconomic Models only}
In the group A models consisting of Model 1 to Model 16, shown in Table \ref{tab1} and \ref{tab2}, we draw an inference of the relationship between Income and Victimization type, given the location. We also delve deeper by analyzing combinations with the other two socioeconomic factors in play. Using the accuracy and 
F1-score metric, we can properly analyze how well our model can generalize future predictions.

\begin{table}[ht]
\centering
{\tiny 
\setstretch{1.0} % Set single spacing
\begin{tabular}{|p{1.0cm}|p{1.4cm}|p{1.7cm}|p{5.5cm}|p{1.2cm}|p{1.2cm}|p{2.2cm}|}
\hline
\textbf{Model} & \textbf{Work Location} & \textbf{IV} & \textbf{Model Formula} & \textbf{Acc} & \textbf{F1 Score} & \textbf{P-Value} \\ \hline
Model 1 & All & Income &   0.284 -0.032*Income & 0.518 & 0.474 & 0.000, 0.000 \\ \hline
Model 2 & Urban & Income &   0.239 -0.033*Income & 0.517 & 0.495 & 0.000, 0.000 \\ \hline
Model 3 & Suburban & Income &   0.392 -0.044*Income & 0.523 & 0.467 & 0.000, 0.000 \\ \hline
Model 4 & Rural & Income &   0.333 -0.023*Income & 0.501 & 0.504 & 0.005, 0.075 \\ \hline
Model 5 & All & Income, Employment &   0.281  -0.033*Income + 0.008*Employment & 0.517 & 0.477 & 0.000, 0.000, 0.586 \\ \hline
Model 6 & Urban & Income, Employment &   0.238  -0.033*Income + 0.005*Employment & 0.517 & 0.495 & 0.000, 0.000, 0.819 \\ \hline
Model 7 & Suburban & Income, Employment &   0.419  -0.041*Income  -0.1*Employment & 0.513 & 0.457 & 0.000, 0.000, 0.004 \\ \hline
Model 8 & Rural & Income, Employment &   0.345  -0.022*Income  -0.027*Employment & 0.518 & 0.507 & 0.004, 0.096, 0.531 \\ \hline
Model 9 & All & Income,  Education &   0.469  -0.027*Income  -0.013* Education & 0.531 & 0.519 & 0.000, 0.000, 0.000 \\ \hline
Model 10 & Urban & Income,  Education &   0.381  -0.028*Income -0.01* Education & 0.527 & 0.51 & 0.000, 0.000, 0.002 \\ \hline
Model 11 & Suburban & Income,  Education &   0.798  -0.036*Income -0.026* Education & 0.552 & 0.549 & 0.000, 0.000, 0.000 \\ \hline
Model 12 & Rural & Income,  Education &   0.396  -0.022*Income  -0.004* Education & 0.516 & 0.499 & 0.017, 0.103, 0.590 \\ \hline
Model 13 & All & Income,  Education, Employment &   0.469  -0.027*Income  -0.013* Education + 0.019*Employment & 0.527 & 0.52 & 0.000, 0.000, 0.000, 0.179 \\ \hline
Model 14 & Urban & Income,  Education, Employment &   0.382  -0.029*Income  -0.01* Education + 0.015*Employment & 0.528 & 0.512 & 0.000, 0.000, 0.002, 0.460 \\ \hline
Model 15 & Suburban & Income,  Education, Employment &   0.785  -0.034*Income -0.024* Education  -0.073*Employment & 0.552 & 0.539 & 0.000, 0.000, 0.000, 0.038 \\ \hline
Model 16 & Rural & Income,  Education, Employment &   0.394  -0.021*Income  -0.004* Education  -0.023*Employment & 0.516 & 0.494 & 0.018, 0.118, 0.671, 0.595 \\ \hline
\end{tabular}
}
\caption{Group A Model Analysis with P-Values, Accuracy and F1-Score}
\label{tab1}
\end{table}

Looking at the first subset (Model 1 - Model 4), the models can be said to be a little bit better than random guessing. The p-values are statistically significant, showing that there must indeed be a relationship happening, however, our model is weak and might not be able to properly generalize it. The second subset (Model 5 - 8) which adds employment does not seem to do as well as the first subset in predicting victimization for all areas except in  rural settings. However, when adding education, the model tends to do better as seen in subset 3 and 4 (Model 9-16). Therefore, in the rural context, using employment as an additional predictor for determining  crime victimization results in better results, whereas, in the cities, we achieve better results when we use education and one can argue from these models that the use of employment in prediction for areas outside rural settings is not useful.

\begin{table}[ht]
\centering
% Set font size and single spacing for the table
{\tiny % Set font size to 10pt
\setstretch{1.0} % Single spacing
\begin{tabular}{|p{0.9cm}|p{1.8cm}|p{1.8cm}|p{2.0cm}|p{2.0cm}|p{2.0cm}|p{1.5cm}|p{2.0cm}|}
\hline
\textbf{Models} & \textbf{Odds Ratios} & \textbf{Standard Errors} & \textbf{CI 2.5\%} & \textbf{CI 97.5\%} & \textbf{-2 Log Likelihood} & \textbf{Sample Size (N)} & \textbf{Significance} \\ \hline
Model 1 & 1.328, 0.968 & 0.036, 0.004 & 0.213, -0.04 & 0.354, -0.025 & -13272.172 & 24000 & ***, *** \\ \hline
Model 2 & 1.27, 0.967 & 0.046, 0.005 & 0.149, -0.043 & 0.329, -0.023 & -7561.847 & 13682 & ***, *** \\ \hline
Model 3 & 1.479, 0.957 & 0.093, 0.009 & 0.21, -0.062 & 0.573, -0.025 & -2417.506 & 4381 & ***, *** \\ \hline
Model 4 & 1.395, 0.977 & 0.117, 0.013 & 0.103, -0.049 & 0.563, 0.002 & -1132.312 & 2053 & **, \\ \hline
Model 5 & 1.325, 0.968, 1.008 & 0.036, 0.004, 0.014 & 0.21, -0.04, -0.02 & 0.352, -0.025, 0.035 & -13272.023 & 24000 & ***, ***, \\ \hline
Model 6 & 1.269, 0.967, 1.005 & 0.046, 0.005, 0.02 & 0.148, -0.043, -0.034 & 0.328, -0.023, 0.043 & -7561.821 & 13682 & ***, ***, \\ \hline
Model 7 & 1.52, 0.96, 0.905 & 0.093, 0.009, 0.035 & 0.236, -0.059, -0.167 & 0.602, -0.023, -0.032 & -2413.355 & 4381 & ***, ***, ** \\ \hline
Model 8 & 1.412, 0.978, 0.973 & 0.119, 0.013, 0.043 & 0.112, -0.048, -0.112 & 0.578, 0.004, 0.058 & -1132.116 & 2053 & **, , \\ \hline
Model 9 & 1.598, 0.974, 0.987 & 0.05, 0.004, 0.002 & 0.37, -0.035, -0.018 & 0.567, -0.019, -0.008 & -13258.003 & 24000 & ***, ***, *** \\ \hline
Model 10 & 1.463, 0.972, 0.99 & 0.065, 0.005, 0.003 & 0.254, -0.039, -0.016 & 0.508, -0.018, -0.004 & -7557.087 & 13682 & ***, ***, ** \\ \hline
Model 11 & 2.222, 0.965, 0.974 & 0.128, 0.009, 0.006 & 0.547, -0.054, -0.037 & 1.049, -0.017, -0.015 & -2406.675 & 4381 & ***, ***, *** \\ \hline
Model 12 & 1.486, 0.979, 0.996 & 0.166, 0.013, 0.008 & 0.071, -0.048, -0.021 & 0.722, 0.004, 0.012 & -1132.167 & 2053 & *, , \\ \hline
Model 13 & 1.599, 0.973, 0.987, 1.019 & 0.05, 0.004, 0.002, 0.014 & 0.371, -0.035, -0.018, -0.009 & 0.568, -0.019, -0.009, 0.046 & -13257.1 & 24000 & ***, ***, ***, \\ \hline
Model 14 & 1.466, 0.972, 0.99, 1.015 & 0.065, 0.005, 0.003, 0.02 & 0.255, -0.039, -0.017, -0.024 & 0.51, -0.019, -0.004, 0.054 & -7556.81 & 13682 & ***, ***, **, \\ \hline
Model 15 & 2.193, 0.966, 0.976, 0.929 & 0.128, 0.009, 0.006, 0.035 & 0.534, -0.053, -0.035, -0.142 & 1.036, -0.016, -0.013, -0.004 & -2404.51 & 4381 & ***, ***, ***, * \\ \hline
Model 16 & 1.483, 0.979, 0.996, 0.977 & 0.166, 0.013, 0.008, 0.044 & 0.069, -0.047, -0.02, -0.11 & 0.72, 0.005, 0.013, 0.063 & -1132.03 & 2053 & *, , , \\ \hline
\end{tabular}
}
\caption{Group A Models Statistical Results}
\label{tab2}
\end{table}

\begin{table}[ht]
\centering
{\tiny 
\setstretch{1.0} % Set single spacing
\begin{tabular}{|p{1.0cm}|p{1.4cm}|p{1.7cm}|p{5.5cm}|p{1.2cm}|p{1.2cm}|p{2.2cm}|}
\hline
\textbf{Model} & \textbf{Work Location} & \textbf{Independent Variables} & \textbf{Model Formula} & \textbf{Acc} & \textbf{F1 Score} & \textbf{P-Value} \\ \hline
Model 17 & All & Income, Age, Marital Status, Gender,  Race &  0.875 -0.016*Income -0.018*Age + 0.073*Marital Status -0.322*Gender -0.087* Race & 0.588 & 0.586 & 0.000, 0.000, 0.000, 0.000, 0.000, 0.000 \\ \hline
Model 18 & Urban & Income, Age, Marital Status, Gender,  Race &  0.722 -0.017*Income -0.015*Age + 0.078*Marital Status -0.332*Gender -0.061* Race & 0.582 & 0.583 & 0.000, 0.002, 0.000, 0.000, 0.000, 0.025 \\ \hline
Model 19 & Suburban & Income, Age, Marital Status, Gender,  Race &  0.993 -0.03*Income -0.015*Age + 0.106*Marital Status -0.436*Gender -0.157* Race & 0.590 & 0.574 & 0.000, 0.002, 0.000, 0.000, 0.000, 0.004 \\ \hline
Model 20 & Rural & Income, Age, Marital Status, Gender,  Race &  1.019 -0.015*Income -0.014*Age + 0.043*Marital Status -0.282*Gender -0.194* Race & 0.582 & 0.576 & 0.000, 0.262, 0.002, 0.197, 0.005, 0.049 \\ \hline
Model 21 & All & Income, Employment, Age, Marital Status, Gender,  Race &  0.879 -0.017*Income + 0.078*Employment -0.019*Age + 0.076*Marital Status -0.32*Gender -0.086* Race & 0.586 & 0.587 & 0.000, 0.000, 0.000, 0.000, 0.000, 0.000, 0.000 \\ \hline
Model 22 & Urban & Income, Employment, Age, Marital Status, Gender,  Race &  0.733 -0.018*Income + 0.071*Employment -0.017*Age + 0.08*Marital Status -0.331*Gender -0.062* Race & 0.590 & 0.592 & 0.000, 0.001, 0.001, 0.000, 0.000, 0.000, 0.022 \\ \hline
Model 23 & Suburban & Income, Employment, Age, Marital Status, Gender,  Race &  0.993 -0.03*Income -0.024*Employment -0.015*Age + 0.105*Marital Status -0.437*Gender -0.158* Race & 0.587 & 0.572 & 0.000, 0.003, 0.500, 0.000, 0.000, 0.000, 0.004 \\ \hline
Model 24 & Rural & Income, Employment, Age, Marital Status, Gender,  Race &  1.022 -0.016*Income + 0.038*Employment -0.014*Age + 0.045*Marital Status -0.282*Gender -0.197* Race & 0.577 & 0.574 & 0.000, 0.232, 0.407, 0.001, 0.173, 0.005, 0.046 \\ \hline
Model 25 & All & Income,  Education, Age, Marital Status, Gender,  Race &  0.943 -0.014*Income -0.006* Education -0.017*Age + 0.074*Marital Status -0.319*Gender -0.086* Race & 0.588 & 0.587 & 0.000, 0.001, 0.015, 0.000, 0.000, 0.000, 0.000 \\ \hline
Model 26 & Urban & Income,  Education, Age, Marital Status, Gender,  Race &  0.775 -0.014*Income -0.005* Education -0.015*Age + 0.079*Marital Status -0.331*Gender -0.06* Race & 0.585 & 0.584 & 0.000, 0.009, 0.146, 0.000, 0.000, 0.000, 0.027 \\ \hline
Model 27 & Suburban & Income,  Education, Age, Marital Status, Gender,  Race &  1.226 -0.025*Income -0.018* Education -0.014*Age + 0.106*Marital Status -0.438*Gender -0.156* Race & 0.607 & 0.590 & 0.000, 0.011, 0.002, 0.000, 0.000, 0.000, 0.004 \\ \hline
Model 28 & Rural & Income,  Education, Age, Marital Status, Gender,  Race &  1.015 -0.015*Income + 0.0* Education -0.014*Age + 0.043*Marital Status -0.282*Gender -0.194* Race & 0.582 & 0.576 & 0.000, 0.269, 0.974, 0.002, 0.196, 0.005, 0.049 \\ \hline
Model 29 & All & Income,  Education, Employment, Age, Marital Status, Gender,  Race &  0.967 -0.014*Income -0.008* Education + 0.084*Employment -0.019*Age + 0.077*Marital Status -0.316*Gender -0.085* Race & 0.584 & 0.583 & 0.000, 0.001, 0.002, 0.000, 0.000, 0.000, 0.000, 0.000 \\ \hline
Model 30 & Urban & Income,  Education, Employment, Age, Marital Status, Gender,  Race &  0.807 -0.015*Income -0.007* Education + 0.077*Employment -0.016*Age + 0.082*Marital Status -0.328*Gender -0.062* Race & 0.591 & 0.592 & 0.000, 0.006, 0.050, 0.000, 0.000, 0.000, 0.000, 0.023 \\ \hline
Model 31 & Suburban & Income,  Education, Employment, Age, Marital Status, Gender,  Race &  1.223 -0.025*Income -0.018* Education -0.009*Employment -0.014*Age + 0.105*Marital Status -0.438*Gender -0.156* Race & 0.608 & 0.591 & 0.000, 0.011, 0.003, 0.812, 0.000, 0.000, 0.000, 0.004 \\ \hline
Model 32 & Rural & Income,  Education, Employment, Age, Marital Status, Gender,  Race &  1.035 -0.016*Income -0.001* Education + 0.039*Employment -0.014*Age + 0.045*Marital Status -0.281*Gender -0.197* Race & 0.574 & 0.572 & 0.000, 0.248, 0.905, 0.403, 0.001, 0.173, 0.006, 0.046 \\ \hline

\end{tabular}
}
\caption{Group B Model Analysis with P-Values, Accuracy and F1-Score}
\label{tab3}
\end{table}

\begin{table}[ht]
\centering
{\tiny % Set font size to 10pt
\setstretch{1.0} % Single spacing
\begin{tabular}{|p{0.9cm}|p{1.8cm}|p{1.8cm}|p{2.0cm}|p{2.0cm}|p{2.0cm}|p{1.5cm}|p{2.0cm}|}
\hline
\textbf{Models} & \textbf{Odds Ratios} & \textbf{Standard Errors} & \textbf{CI 2.5\%} & \textbf{CI 97.5\%} & \textbf{-2 Log Likelihood} & \textbf{Sample Size (N)} & \textbf{Significance} \\ \hline
Model 17 & 2.399, 0.984, 0.982, 1.076, 0.725, 0.917 & 0.076, 0.004, 0.001, 0.009, 0.03, 0.022 & 0.726, -0.024, -0.021, 0.055, -0.38, -0.129 & 1.024, -0.008, -0.015, 0.092, -0.264, -0.045 & -12990.93 & 24000 & ***, ***, ***, ***, ***, *** \\\hline
Model 18 & 2.059, 0.984, 0.985, 1.082, 0.717, 0.941 & 0.1, 0.005, 0.002, 0.012, 0.039, 0.027 & 0.527, -0.027, -0.019, 0.054, -0.409, -0.114 & 0.917, -0.006, -0.012, 0.103, -0.256, -0.007 & -7415.839 & 13682 & ***, **, ***, ***, ***, * \\\hline
Model 19 & 2.699, 0.97, 0.985, 1.112, 0.646, 0.854 & 0.192, 0.01, 0.003, 0.023, 0.07, 0.054 & 0.617, -0.05, -0.022, 0.061, -0.573, -0.264 & 1.369, -0.011, -0.009, 0.15, -0.3, -0.051 & -2342.948 & 4381 & ***, **, ***, ***, ***, ** \\\hline
Model 20 & 2.769, 0.985, 0.986, 1.044, 0.755, 0.823 & 0.261, 0.014, 0.004, 0.033, 0.101, 0.099 & 0.507, -0.042, -0.022, -0.022, -0.48, -0.388 & 1.53, 0.011, -0.005, 0.108, -0.084, -0.001 & -1115.588 & 2053 & ***, , **, , **, * \\\hline
Model 21 & 2.408, 0.983, 1.082, 0.981, 1.079, 0.726, 0.918 & 0.076, 0.004, 0.015, 0.001, 0.009, 0.03, 0.022 & 0.73, -0.025, 0.05, -0.022, 0.057, -0.378, -0.128 & 1.028, -0.01, 0.107, -0.016, 0.094, -0.262, -0.044 & -12976.322 & 24000 & ***, ***, ***, ***, ***, ***, *** \\\hline
Model 22 & 2.082, 0.982, 1.073, 0.983, 1.084, 0.718, 0.94 & 0.1, 0.005, 0.021, 0.002, 0.013, 0.039, 0.027 & 0.538, -0.028, 0.031, -0.02, 0.056, -0.407, -0.115 & 0.929, -0.007, 0.111, -0.013, 0.105, -0.254, -0.009 & -7409.87 & 13682 & ***, ***, ***, ***, ***, ***, * \\\hline
Model 23 & 2.7, 0.971, 0.976, 0.985, 1.111, 0.646, 0.854 & 0.192, 0.01, 0.036, 0.003, 0.023, 0.07, 0.054 & 0.617, -0.049, -0.096, -0.022, 0.06, -0.574, -0.265 & 1.369, -0.011, 0.047, -0.009, 0.149, -0.3, -0.052 & -2342.721 & 4381 & ***, **, , ***, ***, ***, ** \\\hline
Model 24 & 2.778, 0.984, 1.039, 0.986, 1.047, 0.754, 0.821 & 0.261, 0.014, 0.046, 0.004, 0.033, 0.101, 0.099 & 0.51, -0.043, -0.052, -0.023, -0.02, -0.48, -0.391 & 1.534, 0.01, 0.128, -0.006, 0.111, -0.084, -0.003 & -1115.244 & 2053 & ***, , , **, , **, * \\\hline
Model 25 & 2.567, 0.986, 0.994, 0.983, 1.077, 0.727, 0.918 & 0.081, 0.004, 0.002, 0.001, 0.009, 0.03, 0.022 & 0.784, -0.022, -0.011, -0.02, 0.055, -0.377, -0.128 & 1.102, -0.006, -0.001, -0.015, 0.092, -0.261, -0.044 & -12987.98 & 24000 & ***, **, *, ***, ***, ***, *** \\\hline
Model 26 & 2.172, 0.986, 0.995, 0.985, 1.083, 0.718, 0.942 & 0.106, 0.006, 0.003, 0.002, 0.013, 0.039, 0.027 & 0.567, -0.025, -0.011, -0.019, 0.055, -0.407, -0.113 & 0.984, -0.004, 0.002, -0.012, 0.104, -0.254, -0.007 & -7414.781 & 13682 & ***, **, , ***, ***, ***, * \\\hline
Model 27 & 3.407, 0.975, 0.982, 0.986, 1.111, 0.645, 0.856 & 0.206, 0.01, 0.006, 0.003, 0.023, 0.07, 0.054 & 0.821, -0.045, -0.029, -0.02, 0.061, -0.575, -0.263 & 1.631, -0.006, -0.006, -0.008, 0.15, -0.301, -0.049 & -2338.236 & 4381 & ***, *, **, ***, ***, ***, ** \\\hline
Model 28 & 2.759, 0.985, 1.0, 0.986, 1.044, 0.754, 0.823 & 0.284, 0.014, 0.008, 0.004, 0.033, 0.101, 0.099 & 0.459, -0.043, -0.016, -0.022, -0.022, -0.481, -0.388 & 1.571, 0.012, 0.017, -0.005, 0.108, -0.083, -0.001 & -1115.588 & 2053 & ***, , , **, , **, * \\\hline
Model 29 & 2.629, 0.986, 0.992, 1.088, 0.981, 1.08, 0.729, 0.919 & 0.081, 0.004, 0.003, 0.015, 0.001, 0.009, 0.03, 0.022 & 0.807, -0.023, -0.013, 0.055, -0.021, 0.059, -0.374, -0.127 & 1.126, -0.006, -0.003, 0.113, -0.016, 0.096, -0.258, -0.042 & -12971.491 & 24000 & ***, ***, **, ***, ***, ***, ***, *** \\\hline
Model 30 & 2.241, 0.985, 0.993, 1.08, 0.984, 1.085, 0.72, 0.94 & 0.107, 0.006, 0.003, 0.021, 0.002, 0.013, 0.039, 0.027 & 0.598, -0.026, -0.013, 0.036, -0.02, 0.057, -0.405, -0.115 & 1.016, -0.004, 0.0, 0.117, -0.013, 0.106, -0.252, -0.008 & -7407.958 & 13682 & ***, **, , ***, ***, ***, ***, * \\\hline
Model 31 & 3.399, 0.975, 0.982, 0.991, 0.986, 1.111, 0.645, 0.855 & 0.207, 0.01, 0.006, 0.037, 0.003, 0.023, 0.07, 0.054 & 0.818, -0.045, -0.029, -0.081, -0.02, 0.061, -0.575, -0.263 & 1.629, -0.006, -0.006, 0.063, -0.007, 0.15, -0.301, -0.05 & -2338.208 & 4381 & ***, *, **, , ***, ***, ***, ** \\\hline
Model 32 & 2.816, 0.984, 0.999, 1.04, 0.986, 1.046, 0.755, 0.821 & 0.285, 0.014, 0.009, 0.047, 0.004, 0.033, 0.101, 0.099 & 0.477, -0.043, -0.018, -0.052, -0.023, -0.02, -0.48, -0.391 & 1.594, 0.011, 0.016, 0.13, -0.006, 0.111, -0.082, -0.003 & -1115.237 & 2053 & ***, , , , **, , **, * \\\hline

\end{tabular}
}
\caption{Group B Models Statistical Results}
\label{tab4}
\end{table}

\subsubsection{Dependent and Independent Variables}
Diving into the variables, income had a negative coefficient for all group A models as described in Table \ref{tab1}. The odds ratio shown on Table \ref{tab2} were all below 1. These point to the fact that the odds of the event of a violent crime happening when compared to a property crime, given that the person in question is victimized is negatively impacted by income. In other words, the higher you earn, the more potential that you fall victim to a property crime class than a violent crime.

When adding the employment coefficient, we can see from the table that it was all over the place. i.e. in Model 5, employment recorded only a small positive coefficient (+0.008) and a p-value = 0.586, lacking significance, however, in the suburban Model 7, employment was a strong predictor with a slope of -0.1 representing a negative effect and also having a p-value of 0.004 and OR of 0.905. 

Education like income continually showed a negative association with crime victimization. Meaning, higher education levels were more susceptible to property crimes than violent crimes. However, in rural areas, education seems to fail as a solid predictor of crime

\subsubsection{Locational Advantage}
In viewing the differences between the location, it was interesting to see how income consistently showed strong negative associations with the outcome (high statistical significance at ***). Without considering location at all, adding Education or Employment slightly improved the model fit and predictive performance. For example, Model 1 (Income only) achieves Accuracy = 0.518 and F1 = 0.474, while Model 9 (Income + Education) improves Accuracy to 0.531 and F1 to 0.519, and Model 13 (Income + Education + Employment) yields Accuracy = 0.527 and F1 = 0.520.It can be argued that Employment overall does not play a strong role among socio-economic factors in determining the type of crime victimization, when paired with its counterparts.

For urban regions, the statistics is quite similar to the full sample. The addition of Education (Model 10) slightly increases predictive metrics (Accuracy = 0.527, F1 = 0.510) compared to Income alone (Model 2: Accuracy = 0.517, F1 = 0.495). With three predictors (Model 14), Accuracy rises further to 0.528 and F1 to 0.512, which suggests that Education and Employment are important to also consider especially in urban settings.

Overall, in the suburban region, there was a higher record of accuracy and F1-scores than in all the models in 3 out of the 4 cases, with exception of the second subset models, where Income and Employment are the sole predictors. For the baseline on income alone, Model 3 achieves Accuracy = 0.523 and F1 = 0.467. Introducing Education (Model 11) increases Accuracy to 0.552 and F1 to 0.549, representing substantial improvement over the basic model. Similarly, Model 15 (Income, Education, Employment) maintains high performance (Accuracy = 0.552, F1 = 0.539), although slightly lower F1 than Model 11. Here, the inclusion of Education and Employment seems particularly beneficial. In our suburban context, we can argue that the socioeconomic gradients may be more powerful in identifying the outcome.

Rural subsets yielded mixed and less consistent improvements in performance. Model 4 (Income only) has Accuracy = 0.501 and F1 = 0.504. Adding Employment (Model 8) or Education (Model 12) does not markedly improve Accuracy and F1; both remain around 0.516–0.518 for Accuracy and 0.494 –0.507 F1 respectively. When all three predictors are included (Model 16), Accuracy is 0.516 and F1 is 0.494, showing no substantial improvement over simpler models. Moreover, Education and Employment are less consistently significant in the rural subset, which overall reflects current biases.

\subsubsection{Statistical Significance and P-Values}
Table \ref{tab2} shows the p-values per model,corresponding to the intercept and each independent variable as shown in the formula on Table \ref{tab1}. The significance codes (e.g., ***, **, *) indicate the conventional thresholds (where *** p $<$ 0.001, ** p $<$ 0.01, * p $<$ 0.05). Across all models with the exception of rural settings,  Income is highly significant (***), reinforcing the idea of a reliable negative association with the outcome.
Education is also often significant, especially in the All and Urban/ Suburban samples. Its significance is less robust in the Rural sample. Employment’s significance is context-dependent. In some models (e.g., Model 7 Suburban), it is highly significant (***), while in others (e.g., Model 5 All, Model 8 Rural), it is not significant. This variability could be studied in future studies.

\subsubsection{Other Metrics}
Where location was ignored, the sample models (N=24,000) have relatively narrow intervals, which is indicative of our large sample size and leads to a more robust precision. In Suburban and Rural subsets (N=4,381 and N=2,053 respectively) we have a wider confidence intervals (CI), meaning we have less precise estimates. For instance, Model 11 (Suburban) has a relatively wide CI for the intercept (0.547 to 1.049), reflecting fewer observations. Rural models (e.g., Model 16) also show broader CIs and p-values that are less consistently significant, likely due to smaller sample sizes and potentially greater heterogeneity in the data.

Generally the deviance decreases slightly as more predictors are added, indicating better model fit. For example, Model 1 (Income only, All sample) has -2 Log Likelihood = -13272.172, while Model 9 (Income + Education) is -13258.003, and Model 13 (Income + Education + Employment) is -13257.1. The incremental improvements tells us that that while each variable contributes somewhat to explaining variance, no additional variable dramatically changed the model’s overall fit.

\clearpage
\subsection{Group B Models - Socioeconomic Models controlling for Demographic Factors}

\subsubsection{Dependent and Independent Variables (with Demographics)} Diving into the variables now with demographic controls included in Table \ref{tab3}, we see that the addition of demographic factors significantly impact the outcome. Income continues have a negative coefficient across most models, indicating that higher income levels continue to be associated with a lower likelihood of violent crime when compared to property crime, even after accounting for these demographic factors. In other words, controlling for these demographic characteristics does not negate the previously observed negative effect of Income; the higher one earns, the more they are inclined towards the property crime category as the victimized outcome.

When looking at Employment in this new setting (for instance, Models 21 and 22 in Table \ref{tab3}), the coefficients and p-values show some improvement in predictive performance for the All and Urban samples, but remain inconsistent in Suburban and Rural contexts.  In the same context, education still generally remains a significant protective factor against violent crime victimisation, which is a contradiction becauses it therefore increases the chances of being a victim of a property crime. However, in some rural models, education still remains weak as a predictive factor (e.g. Model 28).

Demographic variables themselves, such as Gender and Race, present strong and consistent effects across most models, particularly for the All and Urban samples. Marital Status and Age also appear to adjust the model’s predictive balance, suggesting that demographic controls play a role in shaping the underlying vulnerability profiles.

\subsubsection{Locational Advantage (with Demographics)} In viewing the differences between locations with demographic adjustments, the patterns are quite telling. The All and Urban models (e.g., Model 17 and Model 22) show that adding Age, Marital Status, Gender, and Race helps achieve slightly better Accuracy and F1 scores compared to the initial models without these controls. For example, Model 17 (All) achieves Accuracy = 0.588 and F1 = 0.586, which is higher than previously observed models that relied solely on socio-economic factors. This suggests that demographic elements add some predictive depth, at least for larger or more diverse samples.

In the suburban group, Model 27 stands out, attaining an Accuracy = 0.607 and F1 = 0.590, surpassing earlier models without demographic adjustments. On the other hand, the Rural subset still lags behind in terms of model improvements; Models 20, 24, 28, and 32 show limited gains in predictive metrics, telling us that using these metrics like socioeconomic and demographic might not be enough to predict crime victimization patterns.

\subsubsection{Statistical Significance and P-Values (with Demographics)} Table \ref{tab4} shows the p-values and significance levels for the demographic-controlled models. Similar to the initial set of models, Income retains high statistical significance. Gender and Race frequently emerge as very significant while Age and Marital Status also maintain some significance in many models, especially for All and Urban samples. Employment’s significance remains somewhat context-dependent. For example, in the All and Urban subsets, Employment often attains statistical significance, while in Suburban and Rural contexts, its significance is less stable. Education, when included, is often significant for All and Urban models, but not always for Rural subsets.

\subsubsection{Other Metrics (with Demographics)} The introduction of demographic controls generally helps to improve precision by narrowing the confidence intervals of large samples (e.g., All and Urban). However, in Suburban and especially Rural subsets, the intervalsare still relatively wide, which tells us even though demographic adjustments might be useful, it does not fully compensate for smaller sample sizes or inherent data variability.

Deviance and log-likelihood measures suggest incremental improvements as more demographic variables are added, similar to the trends seen previously. For example, comparing base models that included only Income to those now incorporating Age, Marital Status, Gender, and Race, we see modest decreases in deviance, indicating better model fit. However, as with earlier observations, the addition of demographic factors does not result in a dramatic change in the overall model’s explanatory power. Instead, the enhancements are subtle, indicating that demographic controls refine but do not fundamentally transform the predictive landscape.

\section{Conclusion, Limitations, and Recommendations}

Drawing together the strands of this analysis, the initial hypotheses remain largely supported. Higher income levels are consistently linked to a lower possibility of violent crime victimization, which suggests that as income increases, individuals face a relatively higher likelihood of being victims of property crimes rather than violent crimes. Similarly, higher educational attainment shows a generally protective effect, reducing the risk of violent victimization. Employment status, though less consistent, still shapes victimization patterns in urban locations. 

Demographic factors as well are significant across most models, with men and younger individuals more prone to violent victimization, and certain racial and marital patterns also influencing the odds.

However, there are important limitations to consider. The data source, the NCVS, restricts the range of variables we can incorporate. This constrains our ability to refine models with additional socio-economic or contextual measures, as no external datapoints can be appended to the existing dataset. The observed patterns also vary by geographic subset; while Urban and Suburban contexts often exhibit more robust and stable associations, Rural models yield less consistent improvements, which may be related to smaller sample sizes or nuanced local factors not captured by the available variables.

In light of these constraints, recommendations for further research should focus on what can be accomplished within the existing NCVS framework. Analysts might consider a more granular approach to the current demographic and socio-economic variables to dive deep into subtle patterns. Also, exploring advanced modeling techniques such as mixed-effects models or alternative classification algorithms or even using Machine Learning and AI could help us reason in a different way which help handle the variability within rural data or even improve the current predictive power given the current variables.

%\setstretch{1.0}
\bibliographystyle{plainnat}
\bibliography{reference.bib}

\end{document}